\documentclass[referee]{mn2e}
\usepackage{graphicx}

\newcommand{\refeq}[1]{Equation~(\ref{#1})}
\newcommand{\refsec}[1]{Section~\ref{#1}}
\newcommand{\reffig}[1]{Figure~\ref{#1}}
\newcommand{\reffigs}[2]{Figures~\ref{#1} and~\ref{#2}}

\newcommand{\tobs}{t} % observer time
\newcommand{\tob}{t_0} % observer time
\newcommand{\tjet}{t_\mathrm{jet}} % jet break
\newcommand{\tobsday}{t} % observer time in days
\newcommand{\tjetday}{t_\mathrm{jet}} % jet break in days

\newcommand{\Lobs}{L} % observed luminosity
\newcommand{\thobs}{\Theta} % angle in the lab frame

\newcommand{\thbeam}{\Theta_\mathrm{beam}} % angle of the relativistic beaming
\newcommand{\thb}{\theta_\mathrm{beam}} % angle of the relativistic beaming
\newcommand{\thjet}{\Theta_\mathrm{jet}} % opening angle of the jet

\newcommand{\bB}{\mathbf{B}} % magnetic field (vector)

\def\Dtobs{\tau} % timescale of curvature effect
\def\dtobs{\Delta\tobs} % observed variation timescale

\def\Ejet{E_{\rm jet}}
\def\Eiso{E_{\rm iso}}

\begin{document}

\title[]
 {Is GRB afterglow emission intrinsically anisotropic?}
\author[]{A. M. Beloborodov$^1$\thanks{Also at Astro-Space Center of 
Lebedev Physical Institute, Profsojuznaja 84/32, Moscow 117810, Russia}, 
F. Daigne$^2$\thanks{Institut Universitaire de France}, 
R. Mochkovitch$^2$ and Z. L. Uhm$^3$ \\
$^1$Physics Department and Columbia Astrophysics Laboratory, Columbia
University, New York, NY 10027, USA\\
$^2$Institut d'Astrophysique de Paris, UMR 7095 Universit\'e
Pierre et Marie Curie-Paris 6 - CNRS, \\
98 bis, boulevard Arago, 75014 Paris, France\\
$^3$Institute for the Early Universe and Research Center of MEMS Space Telescope,\\
Ewha Womans University, Seoul 120-750, South Korea\\
\tt e-mail: amb@phys.columbia.edu ; daigne@iap.fr ; mochko@iap.fr ; z.lucas.uhm@gmail.com. }

\maketitle

\begin{abstract}
The curvature of a relativistic blast wave implies that its emission arrives 
to observers with a spread in time. This effect is believed to wash out 
fast variability in the light curves of GRB afterglows. We note that the 
spreading effect is reduced if emission is anisotropic in the 
rest-frame of the blast wave (i.e. if emission is limb-brightened or 
limb-darkened). In particular, synchrotron emission is almost certainly 
anisotropic, and may be strongly anisotropic, depending on details of 
electron acceleration in the blast wave. Anisotropic afterglows can display 
fast and strong variability at high frequencies (above the `fast-cooling' 
frequency). This may explain the existence of bizarre features in the 
X-ray afterglows of GRBs, such as sudden drops and flares. We also note 
that a moderate anisotropy can significantly delay the `jet break' in 
the light curve, which makes it harder to detect.
\end{abstract}

%%%%%%%%%%%%%%%%%%%%%%%%%%%%%%%%%%%%%%%%%%%%%%%%%%%%%%%%%%%%%%%%%%%%%%%%%%
%%%%%%%%%%%%%%%%%%%%%%%%%%%%%%%%%%%%%%%%%%%%%%%%%%%%%%%%%%%%%%%%%%%%%%%%%%
%%%% INTRODUCTION
%%%%%%%%%%%%%%%%%%%%%%%%%%%%%%%%%%%%%%%%%%%%%%%%%%%%%%%%%%%%%%%%%%%%%%%%%%
%%%%%%%%%%%%%%%%%%%%%%%%%%%%%%%%%%%%%%%%%%%%%%%%%%%%%%%%%%%%%%%%%%%%%%%%%%

\section{Introduction}
GRB afterglows are likely produced by relativistic blast waves propagating 
from the center of the explosion. This model is, however, challenged by 
recent observations. In particular, the {\it Swift} satellite revealed 
several puzzling features in the X-ray afterglow. It observed an early 
plateau stage and flares with fast rise and decay times (Nousek et al. 
2006; Burrows et al. 2005).
Less frequent but even more bizarre are sudden drops in the X-ray 
light curve (as steep as $\tobs^{-10}$ in GRB~070110, Troja et al. 2007). 
These behaviors are 
inconsistent with the standard model of afterglow production.

Can the emission from the forward or reverse shock of the blast wave show 
strong variations on timescales $\dtobs\ll\tobs$?
It is usually argued that this is impossible: the spherical 
curvature of the emitting surface (of radius $R$ and Lorentz factor $\Gamma$)
implies a spread in arrival times of its emission, which washes out 
variability on timescales shorter than
\begin{equation}
\label{eq:Dtobs}
    \Dtobs = \frac{R}{2c\Gamma^2}.
\end{equation}
For a relativistic blast wave, this duration is comparable to the observed 
time passed since the beginning of the explosion, $\Dtobs\sim\tobs$. This 
appears to prohibit any rapid and strong variations in the light curve (see 
Ioka et al. 2005 for 
discussion).

Therefore, the observed fast variability in afterglows is usually 
associated with additional emission from radii much smaller than the 
blast-wave radius. This model invokes a late activity of the central 
engine (Zhang et al. 2006). The material ejected at large $\tobs$ and 
emitting at radii 
$R\ll\Gamma^2\tobs c$ will have $\Dtobs\ll\tobs$ and can produce flares 
with $\dtobs\ll\tobs$. Note however that (i) it is unclear in this model 
why the observed flares have the approximately universal 
$\dtobs/\tobs\sim 0.1$ (Chincarini et al. 2007; Lazzati \& Perna 2007), 
(ii) the very steep drops at the end of some plateaus can hardly be 
explained by this model unless it assumes that the entire plateau is 
produced at small radii inside the ejecta and the emission from the blast 
wave is negligible (Kumar, Narayan \& Johnson 2008).

Another difficulty for GRB theory is that many afterglows lack the 
predicted `jet breaks' (Burrows \& Racusin 2006; Sato et al. 2007): 
only a small fraction of 
afterglow light curves 
show a clear achromatic break that is expected from jets 
(Willingale et al. 2007).\footnote{Many afterglows show chromatic breaks, 
which occur either in the X-ray or in the optical, but not in both bands.} 
Some bursts show X-ray light curves extending for tens to hundreds of days 
with a constant temporal slope (Grupe et al. 2007). 
The interpretation of these observations is difficult and often leads one 
to assume large jet opening angles, implying in some cases 
extremely high energy
for the explosion (Shady et al. 2007). 

An implicit assumption in the general discussion of these puzzling features
is that the emission is isotropic in the rest frame of the relativistically 
moving 
source (however, see Lyutikov 2006).
In this paper, we discuss the effects of a possible 
anisotropy and suggest that they can help explain observations.
In \refsec{sec:response} we 
write down
a general formula for the observed
flux from a flashing sphere when the emission is anisotropic in the source
rest frame. In \refsec{sec:effects} we list
the consequences of anisotropy for the curvature effect,
the jet break in the afterglow light curve and the size of the radio image of 
the blast wave. In \refsec{sec:meca} we 
consider the standard radiative mechanism of afterglows -- synchrotron 
emission -- and discuss its anisotropy.
The results are summarized in \refsec{sec:discussion}.

%%%%%%%%%%%%%%%%%%%%%%%%%%%%%%%%%%%%%%%%%%%%%%%%%%%%%%%%%%%%%%%%%%%%%%%%%%
%%%%%%%%%%%%%%%%%%%%%%%%%%%%%%%%%%%%%%%%%%%%%%%%%%%%%%%%%%%%%%%%%%%%%%%%%%
%%%% RESPONSE FUNCTION OF A FLASHING SPHERE
%%%%%%%%%%%%%%%%%%%%%%%%%%%%%%%%%%%%%%%%%%%%%%%%%%%%%%%%%%%%%%%%%%%%%%%%%%
%%%%%%%%%%%%%%%%%%%%%%%%%%%%%%%%%%%%%%%%%%%%%%%%%%%%%%%%%%%%%%%%%%%%%%%%%%

\section{Lightcurve from a flashing sphere}
\label{sec:response}
Let an energy $E$ (measured in the lab frame) be instantaneously emitted by 
a sphere of radius $R$, which is expanding with a Lorentz factor $\Gamma$ 
and a velocity $\beta=v/c=\left(1-\Gamma^{-2}\right)^{1/2}$.  
Distant observers will see the emitted radiation 
extend over a range of 
arrival times $\tobs$ due to the curvature of the emitting surface.
The observed light curve $\Lobs(\tobs)$
from the flashing sphere can be thought of as a Green function 
of afterglow emission or `response function'
that describes the curvature effect.
It depends on the intrinsic angular distribution of the source intensity.
Let $\theta$ be the photon angle with respect to the radial direction
in the local rest frame of the emitting sphere. The intrinsic (comoving) 
angular distribution of emission (per unit solid angle) can be described by 
function $A(\theta)$ normalized by 
\begin{equation} 
  \int A(\theta)\,d\Omega=4\pi.
\end{equation}
Isotropic emission in the comoving frame corresponds to $A(\theta)=1$.
The photon angle in the static lab frame $\thobs$ is related to $\theta$ by
\begin{equation}
  \cos\thobs=\frac{\cos{\theta}+\beta}{1+\beta\cos{\theta}}\ .
  \label{eq:thlab}
\end{equation}

A distant observer will first see photons emitted along the line of sight
with $\theta=\thobs=0$. Let $\tob$ be the arrival time of these first photons.
Photons received at a later time $\tobs$ come from a larger co-latitude
$\thobs$ on the emitting sphere, related to $\tobs$ by
\begin{equation}
  \tobs-\tob=\frac{R}{c}\left(1-\cos\thobs\right).
\end{equation}
A time interval $\delta\tobs$ corresponds to a ring 
$\delta\cos\thobs=c\delta\tobs/R$ on the sphere.
The true energy emitted by this ring is 
$\delta E={1\over 2} E\,\delta\cos\thobs$.
The ring is viewed at angle $\Theta$ with respect to its normal, and 
the observed photons are Doppler-boosted in energy by the factor 
${\cal D}=\Gamma^{-1}(1-\beta\cos\Theta)^{-1}$. Doppler effect also 
compresses the solid angle of emission by the factor of ${\cal D}^{-2}$. 
Together with anisotropy $A(\theta)$, Doppler effect determines the 
amplification factor for the {\it apparent} isotropic equivalent of 
emitted energy, 
\begin{equation}
   \delta E_\mathrm{app}=A(\theta)\,{\cal D}^3\,\frac{\delta E}{\Gamma}.
\end{equation}
The observed luminosity is $\Lobs=\delta E_\mathrm{app}/\delta\tobs$, which 
yields
\begin{equation}
   \Lobs=\frac{cE}{2R}\frac{A(\theta)}{\Gamma^4(1-\beta\cos\thobs)^3},
\end{equation}
  where $\theta$ and $\thobs$ are related by \refeq{eq:thlab}. 
Substitution of $\cos\thobs(\tobs)=1-c\left(\tobs-\tob\right)/R$ and 
$\theta[\thobs(\tobs)]$ gives an explicit expression for the light curve 
produced by the flashing sphere. 
These equations simplify in the limit $\Gamma\gg 1$,
\begin{equation}
\label{eq:th}
  \cos\theta=\frac{\Dtobs-(\tobs-\tob)}{\Dtobs+(\tobs-\tob)}, 
\end{equation}
\begin{equation}
\label{eq:rf}
   \Lobs=\frac{2E}{\Dtobs}\,A(\theta)
     \left(1+\frac{\tobs-\tob}{\Dtobs}\right)^{-3},
\end{equation}
where $\Dtobs$ is defined in \refeq{eq:Dtobs}.

%%%%%%%%%%%%%%%%%%%%%%%%%%%%%%%%%%%%%%%%%%%%%%%%%%%%%%%%%%%%%%%%%%%%%%%%%%
%%%%%%%%%%%%%%%%%%%%%%%%%%%%%%%%%%%%%%%%%%%%%%%%%%%%%%%%%%%%%%%%%%%%%%%%%%
%%%% POSSIBLE CONSEQUENCES
%%%%%%%%%%%%%%%%%%%%%%%%%%%%%%%%%%%%%%%%%%%%%%%%%%%%%%%%%%%%%%%%%%%%%%%%%%
%%%%%%%%%%%%%%%%%%%%%%%%%%%%%%%%%%%%%%%%%%%%%%%%%%%%%%%%%%%%%%%%%%%%%%%%%%

\section{Some consequences of anisotropy}
\label{sec:effects}
An isotropic source, $A(\theta)=1$, after Doppler transformation to the 
static frame emits 75 per cent of the energy within $\thbeam=1/\Gamma$.
Let us now consider an anisotropic source, $A(\theta)\neq 1$.
We will assume that the anisotropy 
has a front-back symmetry,
$A(\theta)=A(\pi-\theta)$;
then the net momentum of emitted photons vanishes in the source frame.

Consider, for instance, `limb-darkened' emission, which is weak near 
$\theta=\pi/2$ and strong near $\theta=0,\pi$. Doppler transformation 
to the lab frame strongly amplifies the radiation with $\theta\approx 0$ 
and weakens radiation with $\theta\approx \pi$.
As a result, a bright narrow beam is created,
so that 75 per cent of energy is now 
concentrated within $\thbeam=(k \Gamma)^{-1}$.
Here $k>1$ is a measure of the enhanced beaming of radiation in the lab 
frame. The beam $\thobs<\thbeam$ is emitted with $\theta<\thb$ in 
the source frame, and one can show that $\thb$ is related to $k$ by
\begin{equation}
   k \approx \sqrt{\frac{1+\cos{\thb}}{1-\cos{\thb}}}\ . 
\end{equation}
The increased beaming in the lab frame ($k>1$) due to limb-darkening in 
the source frame ($\cos\thb>0$) has several observational consequences 
that we list below.

\subsection{Curvature effect}

The curvature effect is expected to control the observed light curve
if the source power suddenly drops.
The observed luminosity $L(t)$ responds to the drop with a 
delay according to \refeq{eq:rf}. If the emission is isotropic in the 
source frame, the delay timescale is $\tau\sim t$, and the steepest 
possible decay is $L(t)\propto t^{-\alpha}$ with $\alpha=3$
(e.g. Kumar \& Panaitescu 2000). Since limb-darkening of the source 
implies stronger beaming, $\thbeam=(k\Gamma)^{-1}$ 
instead of $\Gamma^{-1}$, most of the energy 
is radiated on a shorter timescale, $\Dtobs/k^2$, and 
the slope $\alpha$ is much steeper.

We will discuss this effect in more detail in Section~4.
It turns out that a similar conclusion holds for the opposite, 
limb-brightened, type of anisotropy, when emission is 
strong near $\theta=\pi/2$ and weak near $\theta=0,\pi$.

\subsection{Jet break}

GRB jets are likely to have a small opening angle $\thjet\ll 1$,
which reduces their energy $\Ejet\approx(\thjet^2/2)\Eiso$ 
(here $\Eiso$ is the isotropic equivalent of the jet kinetic energy).
A break should be observed in the afterglow light curve at moment $\tjet$ 
when the relativistic beaming angle $\thbeam$ becomes larger than 
$\thjet$ (Rhoads 1997), and the value of $\thjet$ may be inferred from 
observed $\tjet$. If the afterglow source is limb-darkened so that 
$\thbeam=(k\Gamma)^{-1}$, the jet-break condition becomes
$\Gamma\approx (k\thjet)^{-1}$, i.e. effectively $\thjet$ is replaced
by $k\thjet$. The standard light-curve analysis can only give the value 
of $k\thjet$, which overestimates the true $\thjet$ by the factor of $k$.
The true $\Ejet$ for a limb-darkened jet is reduced by the factor of 
$k^{-2}$ compared with the usual estimate. 

Limb-darkening also implies a significant delay in $\tjet$. For example,
consider a blast wave decelerating in a uniform medium. Its Lorentz factor 
decreases as $\Gamma\propto t^{-3/8}$. The jet break occurs
when $\Gamma\approx (k\thjet)^{-1}$, and this moment is delayed by the 
factor of $k^{8/3}$. The usual expression for $\tjet$ then becomes
  \begin{equation}
     \tjet\approx k^{8/3}\left(\frac{E_{\rm iso,53}}{n}\right)^{1/3} 
     \left(\frac{\thjet}{0.1}\right)^{8/3} \mathrm{d}.
  \end{equation}
Similarly, for a blast wave decelerating in a 
wind medium, $\Gamma\propto t^{-1/4}$ and hence $\tjet\propto k^4$.
Even a moderate anisotropy (e.g. $k = \sqrt{3}$, which corresponds to 
limb-darkening with $\thb\sim 60^{\rm o}$ in the source frame) can delay 
the jet break by a large factor ($\sim 4.3$ in a uniform medium, $\sim 9$ 
in a wind). This could be enough to not detect the jet break with current 
observational capabilities as the afterglow is dim at late times and the 
spectral coverage is incomplete to test achromaticity.

  \subsection {Apparent size of the radio afterglow source}

VLBI observations 
  provided
the angular size of the radio image of a few GRB 
afterglows (Frail et al. 1997; Taylor et al. 2005) which helps constrain 
the ratio of the blast wave energy to the density of the environment.
The best data have been obtained for GRB~030329 and seem to favor a 
blast wave in a uniform medium with 
$\Ejet/n \sim (1-5)\times 10^{51}$~erg~cm$^3$ 
and $\thjet\sim 0.1$~rad (Pihlstr\"om et al. 2007).

The apparent size of the afterglow source is given by 
(e.g. Oren, Nakar \& Piran 2004) 
\begin{equation}
   R_\mathrm{\perp}\simeq R\,\thobs,
  \qquad \thobs=\min{\left(\thbeam; \thjet \right)},
\end{equation}
where $R$ is the radius of the emitting shell.  With increased beaming 
due to limb-darkening, $\thbeam=(k\Gamma)^{-1}$ and $R_\perp$ is reduced
by the factor $k^{-1}$. For a blast wave in a uniform medium, a derivation 
similar to that in Oren et al. (2004) gives the relation between 
the ratio of the true jet energy to the external density and 
observed $R_\perp$, $t$ and $\tjet$,
\begin{equation}
   \frac{\Ejet}{n}\approx 10^{51}\,k^4
       \left(\frac{R_\perp}{6\times 10^{16}\rm cm}\right)^6 
       \times \left\lbrace\begin{array}{cl} 
    \tjetday^{3/4}\, \tobsday^{-15/4} & \mathrm{(before\ jet\ break)}\\
    \tobsday^{-3} & \mathrm{(after\ jet\ break)}\\
\end{array}\right. \ \mathrm{erg~cm}^3
\label{eq:radio1}
\end{equation}
One can show that $k$ drops out from the similar relation derived for 
blast waves in wind media, i.e. in that case limb-darkening does 
not affect the relation.

%%%%%%%%%%%%%%%%%%%%%%%%%%%%%%%%%%%%%%%%%%%%%%%%%%%%%%%%%%%%%%%%%%%%%%%%%%
%%%%%%%%%%%%%%%%%%%%%%%%%%%%%%%%%%%%%%%%%%%%%%%%%%%%%%%%%%%%%%%%%%%%%%%%%%
%%%% ANISOTROPY OF SYNCHROTRON EMISSION
%%%%%%%%%%%%%%%%%%%%%%%%%%%%%%%%%%%%%%%%%%%%%%%%%%%%%%%%%%%%%%%%%%%%%%%%%%
%%%%%%%%%%%%%%%%%%%%%%%%%%%%%%%%%%%%%%%%%%%%%%%%%%%%%%%%%%%%%%%%%%%%%%%%%%
  
\section{Anisotropy of synchrotron emission}
\label{sec:meca}

Afterglow is commonly interpreted as synchrotron emission. Its anisotropy
naturally results from a preferred orientation of the magnetic field $\bB$.
Magnetic fields inside GRB jets are generally expected to be transverse to 
the jet direction, as radial expansion quickly suppresses the longitudinal 
component. Internal or external shocks can generate magnetic fields only in 
the shock plane (e.g. Medvedev \& Loeb 1999).
Thus, in various models of afterglow production\footnote{
     At present, the origin of afterglow emission is unclear. 
     The standard forward-shock model is in conflict with data, and it is 
     possible that the afterglow is produced by a long-lived reverse shock
     (Uhm \& Beloborodov, 2007; Genet, Daigne \& Mochkovitch, 2007).}
it is reasonable to suppose that the magnetic field in the source is 
perpendicular to its velocity, $\bB=\bB_\perp$. In addition, we assume that 
$\bB_\perp$ is tangled on a scale much smaller than $R/\Gamma$, so that a 
distant observer will see a superposition of emissions from many 
domains with random orientations of $\bB_\perp$. 
This assumption is motivated by the low polarization in observed afterglows,
typically less than a few per cent (e.g. Covino et al. 1999). 

The anisotropy of synchrotron emission may be expected to be 
moderate
if the emitting electrons have an isotropic distribution 
(see e.g.
calculations 
by Granot et al. (1999) for three possible geometries of the magnetic field). 
Even in this case, anisotropy is present because 
$\bB$ is confined to a plane. After averaging over random directions of 
$\bB_\perp$, one finds the angular distribution of emitted power per electron,
\begin{equation}
\label{eq:iso}
  \frac{dP}{d\Omega}=\frac{P_0}{4\pi}\,A_0(\theta), \qquad
\mathrm{with}\,  \ \   A_0(\theta)=\frac{3}{4}\,(1+\cos^2{\theta}).
\end{equation}
Here 
$P_0=(\sigma_\mathrm{T} c / 6\pi ) \gamma^2 B^2$ 
is the power of synchrotron emission per electron and $\sigma_{\rm T}$ is 
Thomson cross section. The resulting radiation is limb-darkened.

In reality, the electron distribution may not be isotropic:
electrons may be preferentially accelerated along or perpendicular to 
the magnetic field, depending on the acceleration mechanism.
For instance, the details of electron acceleration in relativistic shocks 
  remain uncertain, despite significant progress in numerical simulations 
  (e.g. Hededal et al. 2004; Spitkovski 2008; Nishikawa et al. 2009),
  and other mechanisms are possible. We therefore consider both types of 
  electron anisotropy.

Let $\alpha$ be the pitch-angle of an electron with respect to the 
magnetic field. We will describe the distribution of electron directions 
$\mathbf{\Omega}_e$ by the function $f(\alpha)$, normalized by 
$\int f(\alpha)\,d\Omega_e=4\pi$. 
Synchrotron emission from each electron is strongly beamed along its 
velocity, and together the electrons emit radiation with angular 
distribution,
\begin{equation}
\label{eq:aniso}
  \frac{dP}{d\Omega_e}=\frac{3P_0}{8\pi}\,\sin^2{\alpha}\, f(\alpha).
\end{equation}
The average power per electron is now given by, 
\begin{equation}
\label{eq:eta}
    P= \eta P_0, \qquad
    \eta\equiv \frac{3}{8\pi}\int \sin^2{\alpha}~f(\alpha)\,d\Omega_e.
\end{equation}

\refeq{eq:aniso} describes the angular distribution relative to the local 
magnetic field. The angle $\alpha$ between the magnetic field and the 
observer's line of sight is given by $\cos\alpha=\sin\theta\cos\phi$ 
where $0\leq\phi<2\pi$ depends on the orientation of $\bB=\bB_\perp$. 
Using \refeq{eq:aniso} and averaging over random directions of $\bB_\perp$, 
one finds that the synchrotron emission has the following angular distribution, 
\begin{equation}
\label{eq:A_theta}
  \frac{dP}{d\Omega}=\frac{\eta P_0}{4\pi}\, A(\theta), 
  \,\,\, \mathrm{with}\,\,\,
  A(\theta)=\frac{3}{2\eta}\,\frac{1}{2\pi}\int_0^{2\pi}
   \left(1-\sin^2{\theta}\cos^2{\phi}\right)\,f
   \left[\arccos(\sin{\theta}\cos{\phi})\right]\,d\phi.
\end{equation}

Let us consider two toy models:
\begin{equation}
   f_1(\alpha)\propto\left(a^2+\sin^2\alpha\right)^{-3}
\qquad
\mbox{and}
\qquad
   f_2(\alpha)\propto\left(a^2+\cos^2\alpha\right)^{-3},
\end{equation}
where $a$ defines a characteristic 
  beaming angle of the electron distribution. 
These expressions represent two opposite cases where electrons are 
preferentially accelerated along $\bB$ (distribution $f_1$) and 
perpendicular to $\bB$ (distribution $f_2$).
Both distributions becomes isotropic if $a\gg 1$. The opposite limit 
$a\ll 1$ describes the maximum possible anisotropy:
$f_1=\delta(\alpha)$ and $f_2=\delta(\alpha-\pi/2)$. 
By varying the parameter $a$ from $\infty$ to $0$, one can explore 
the effect of increasing electron anisotropy on synchrotron emission.

\begin{figure}
\begin{center}
\begin{tabular}{cc}
\includegraphics[width=0.45\textwidth]{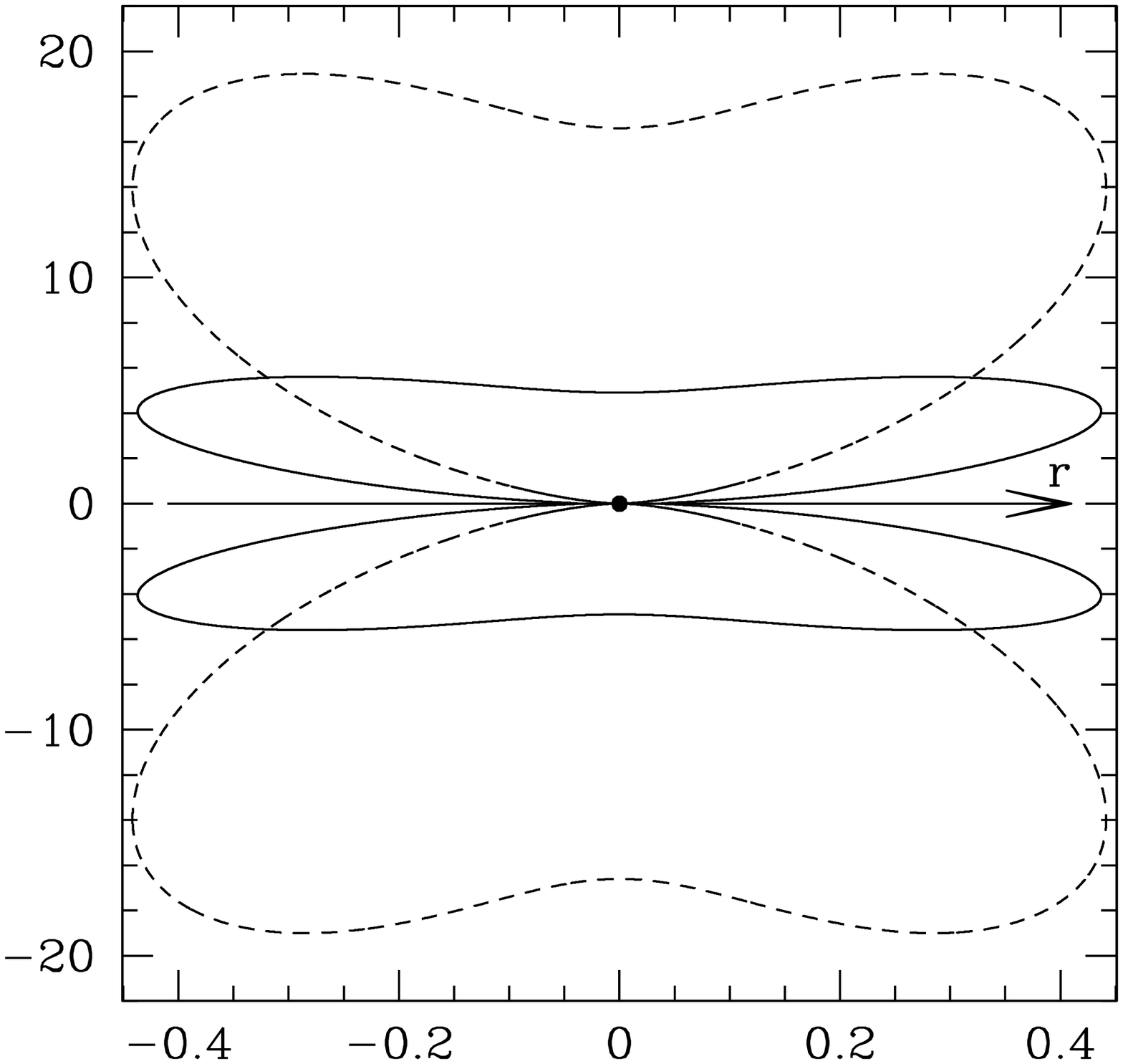} &
\includegraphics[width=0.45\textwidth]{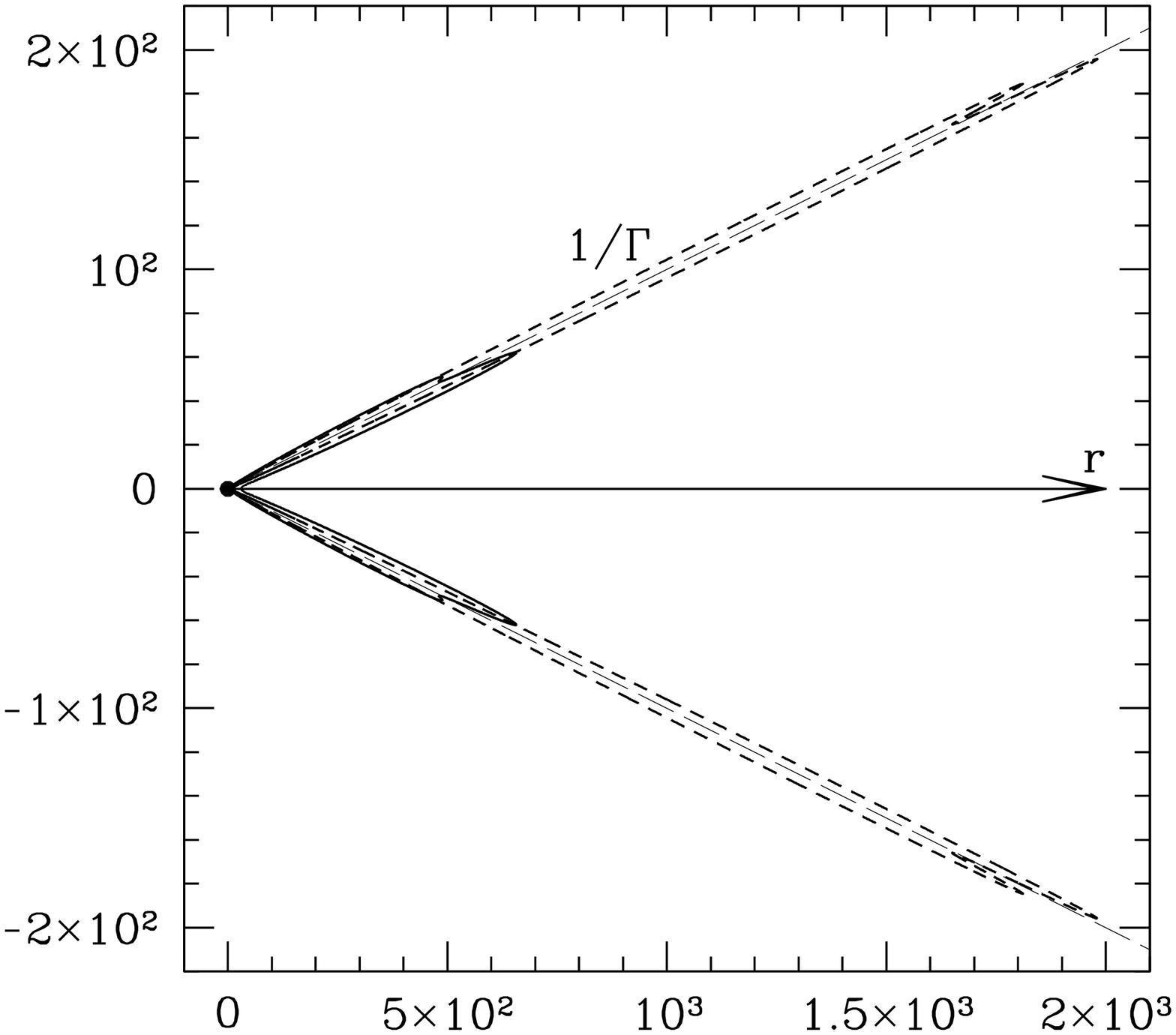} \\
\end{tabular}
\end{center}
\caption{
Diagram of angular distribution of radiation $A_1(\theta)$ (electrons 
accelerated 
  preferentially
along $\bB$) measured in the 
  source rest frame
(left) and transformed to the observer frame using $\Gamma=10$ (right). 
Solid curves correspond to $a=0.1$ and dashed to $a=0.03$. Long-dashed lines in the right panel show the cone of opening angle $1/\Gamma$.
Rotation of the shown curve about the 
  horizontal
axis gives the 3-dimensional diagram. 
}      
\label{fig:A1}
\end{figure}

The angular distributions $A_1(\theta)$ and $A_2(\theta)$ produced by $f_1$ 
and $f_2$ are shown in \reffigs{fig:A1}{fig:A2}. The plot of $A_1(\theta)$ 
resembles a butterfly with a half-angle $\sim a$ around $\theta=\pi/2$, 
i.e. the source is limb-brightened. Therefore, the angular distribution of 
radiation in the fixed lab frame has a sharp peak at $\thobs=\Gamma^{-1}$ 
when $a\ll 1$. By contrast, $A_2(\theta)$ is concentrated near 
$\theta=0,\pi$ with a half-angle $\sim a$; 
the resulting limb-darkening remains, however, finite even if 
$a\rightarrow 0$. In this limit, one finds
\begin{equation}
   A_2(\theta)=\frac{2}{\pi\sin\theta} \qquad (a\rightarrow 0).
\end{equation}
The source is weakened at $\theta=\pi/2$ by the modest factor of $2/\pi$ 
compared with isotropic emission.\footnote{
    Electron distribution with $f_2=\delta(\alpha-\pi/2)$ is a disk in 
    momentum space, with the axis along $\bB$. After averaging over 
    random orientations of the disk axis in the transverse plane (random 
    directions of $\bB$) a strong anisotropy is found in $A_2(\theta)$, with 
    a sharp peak in the longitudinal directions $\theta=0,\pi$. However, a 
    significant `wing' of emission remains present at large $\theta\sim\pi/2$.
}
Note also the difference in the emission power $P$ for the two 
distributions. For $a\ll 1$ one obtains
\begin{equation}
\label{eq:eta12}
    \eta_1\approx \frac{3}{2}\,a^2, \qquad \eta_2\approx \frac{3}{2}.
\end{equation}

%%%%%%%%%%%%%%%%%%%%%%%%%%%%%%%%%%%%%%%%%%%%%%%%%%%%%%%%%%%%%%%%%%%%%%%
%%%%%%%%%%%%%%%%%%%%%%%%%%%%%%%%%%%%%%%%%%%%%%%%%%%%%%%%%%%%%%%%%%%%%%%

%%%%%%%%%%%%%%%%%%%%%%%%%%%%%%%%%%%%%%%%%%%%%%%%%%%%%%%%%%%%%%%%%%%%%%%
\begin{figure} 
\begin{center}
\begin{tabular}{cc}
\includegraphics[width=0.45\textwidth]{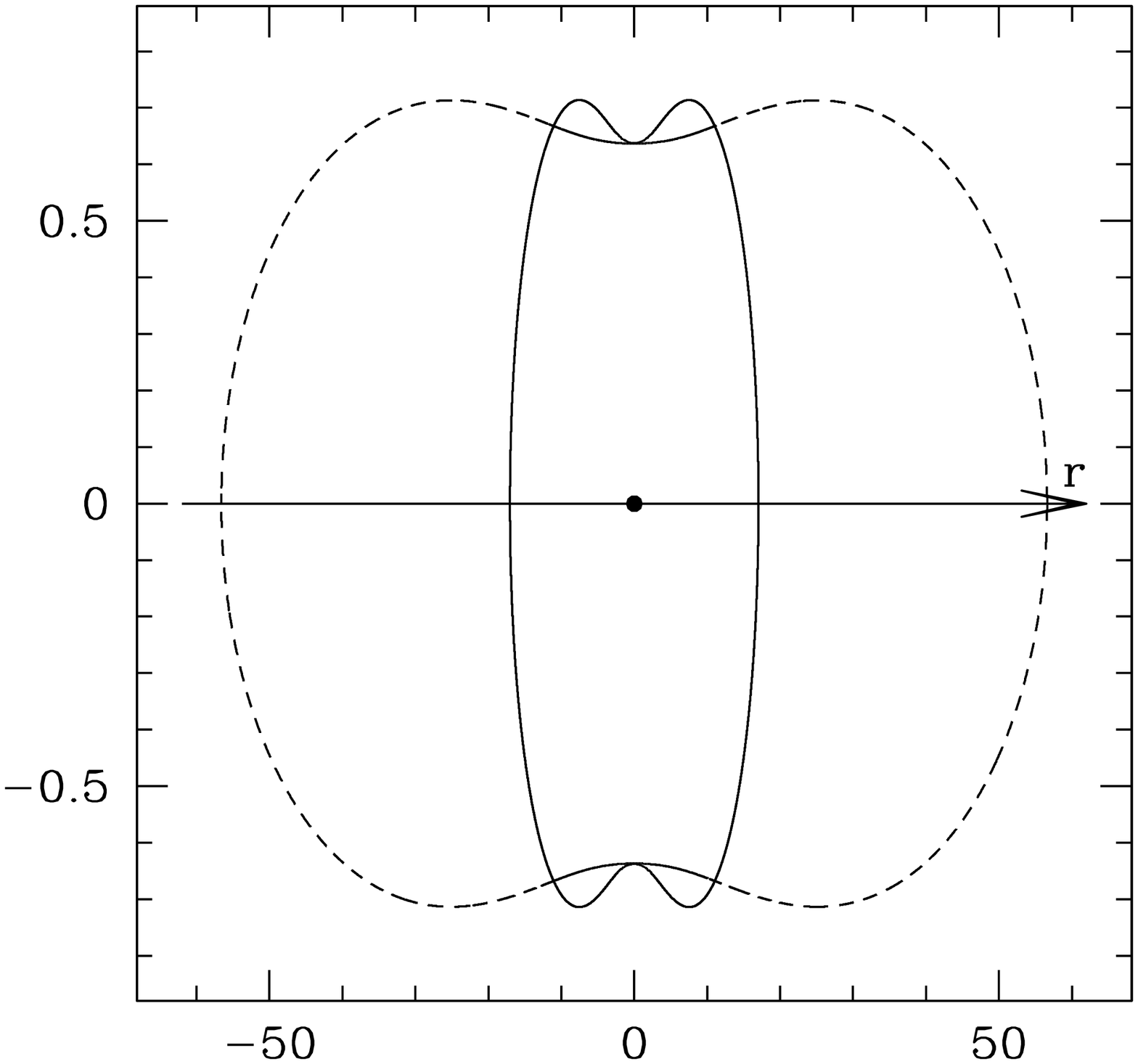} &
\includegraphics[width=0.45\textwidth]{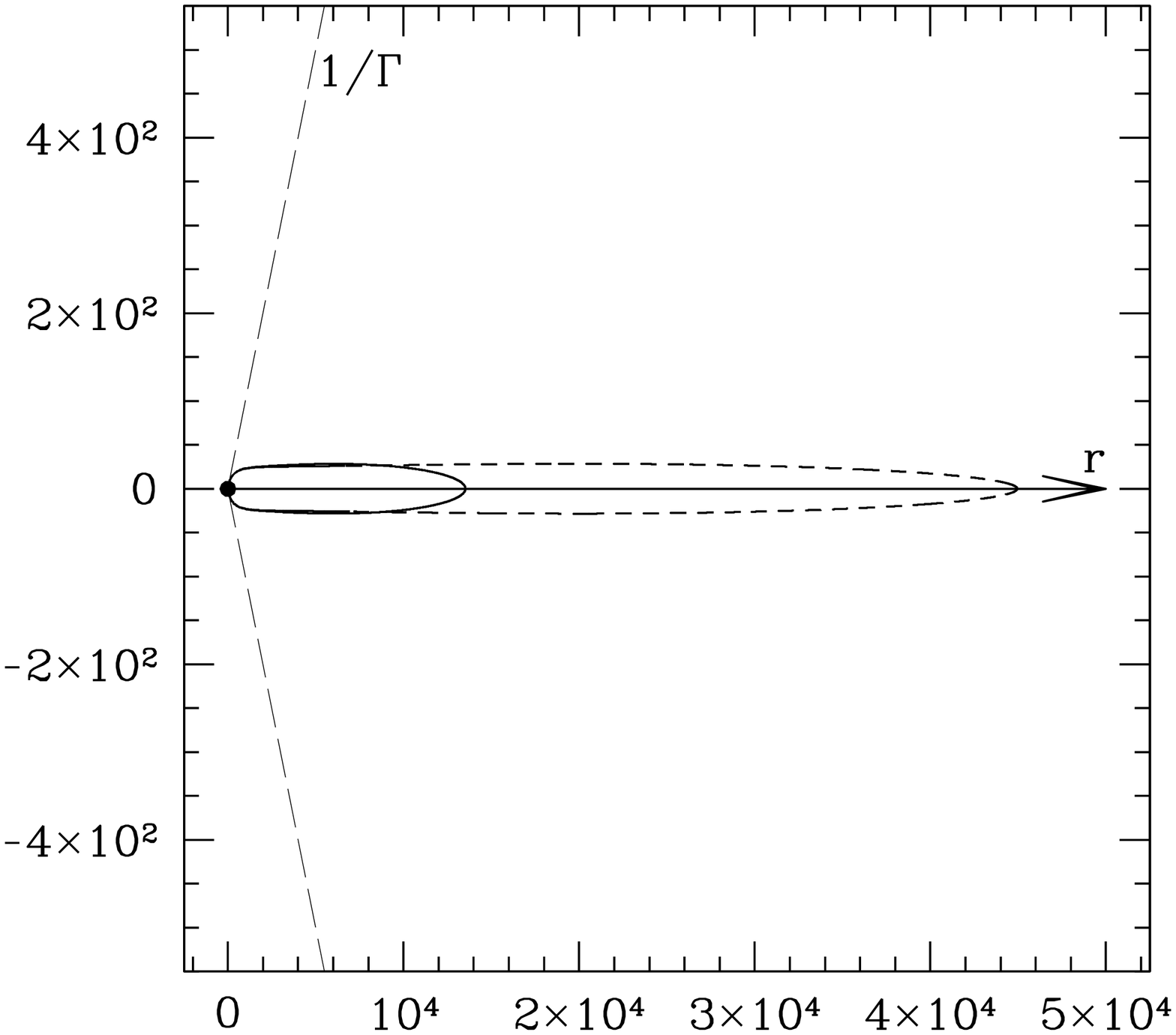} \\
\end{tabular}
\end{center}
\caption{Same as in \reffig{fig:A1} but for the angular distribution 
$A_2(\theta)$ (electrons accelerated
  preferentially
 perpendicular to $\bB$).}
\label{fig:A2}
\end{figure} 
%%%%%%%%%%%%%%%%%%%%%%%%%%%%%%%%%%%%%%%%%%%%%%%%%%%%%%%%%%%%%%%%%%%%%%%

%%%%%%%%%%%%%%%%%%%%%%%%%%%%%%%%%%%%%%%%%%%%%%%%%%%%%%%%%%%%%%%%%%%%%%%
\begin{figure} 
\centerline{\includegraphics[width=0.5\textwidth]{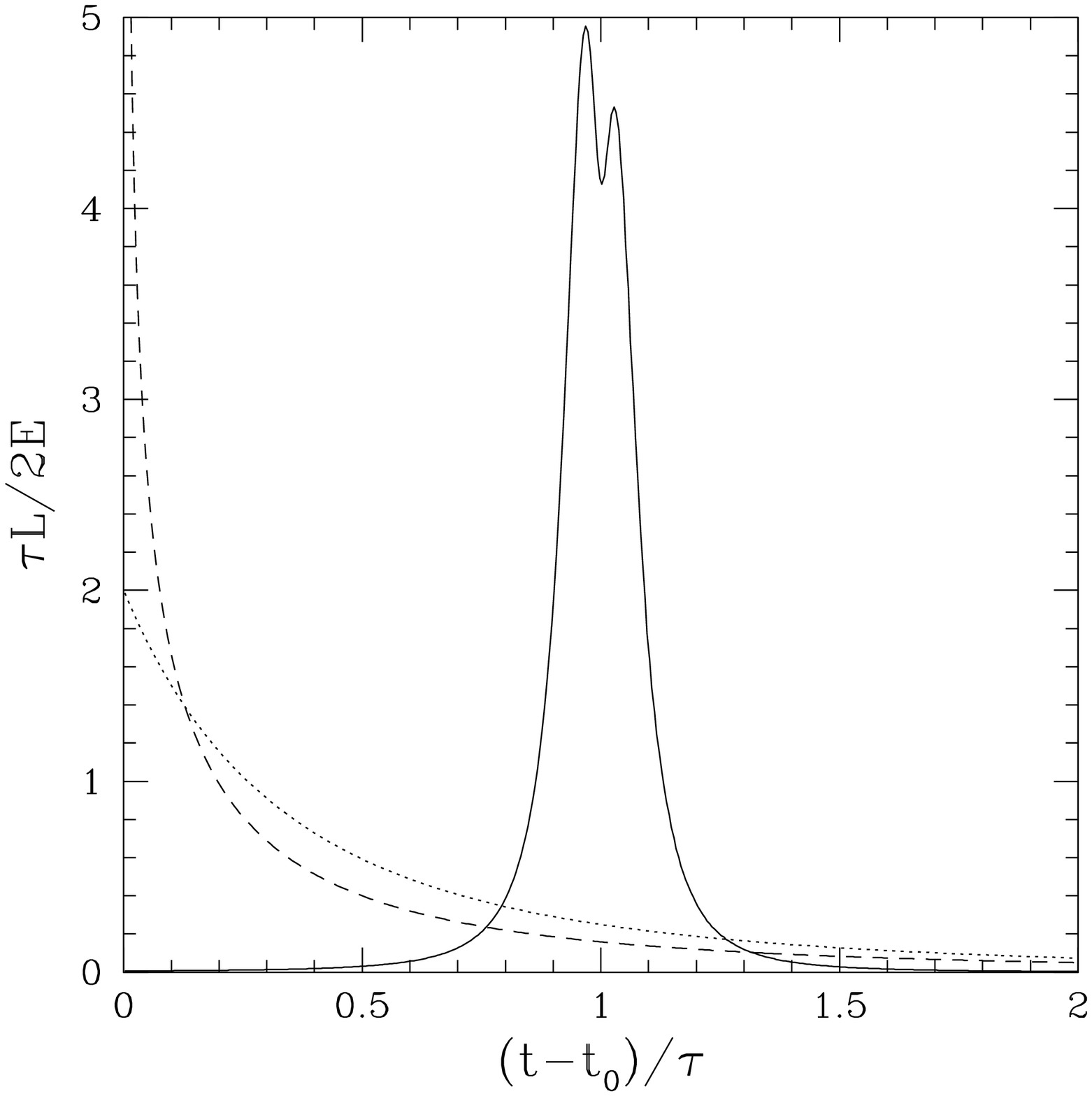}} 
\caption{
  Lightcurve from
a flashing sphere with three different 
angular distributions of emission in the plasma rest frame:
isotropic (dotted curve), $A_1(\theta)$ with $a=0.03$ (solid curve),
and $A_2(\theta)$ with $a=0.03$ (dashed curve). The units for $\tobs$
and $\Lobs$ are indicated on the axes ($\Dtobs=R/2\Gamma^2c$); with 
these units the area under each curve is unity.
 The first photon from the flashing sphere reaches the observer at $\tobs=t_0$. 
 The dashed curve has the maximum of about $13$ at $\tobs=t_0$.
}
\label{fig:response}
\end{figure} 
%%%%%%%%%%%%%%%%%%%%%%%%%%%%%%%%%%%%%%%%%%%%%%%%%%%%%%%%%%%%%%%%%%%%%%%

The anisotropy of emission $A(\theta)\neq 1$ can have a strong impact
on the afterglow light curves as discussed in section~3. In particular,
the curvature effect, which is described by the 
light curve from a flashing sphere (the response function, \refeq{eq:rf}), 
is changed. \reffig{fig:response} shows 
the response function for $A_1(\theta)$ and $A_2(\theta)$ with $a=0.03$.
Compared to the isotropic case, $A(\theta)=1$, the emitted pulse becomes 
very narrow if $A(\theta)=A_1(\theta)$, i.e. for the model where electrons 
are preferentially accelerated along the magnetic field. Photon arrival times 
then 
concentrate
near a particular $t-t_0\approx\tau$ (which corresponds to a 
particular $\thobs=\Gamma^{-1}$) because the limb-brightened radiation is 
mainly emitted near $\theta=\pi/2$. In the model with angular distribution 
$A_2(\theta)$, the profile of the response function is steeper than in the 
isotropic case, but can never be as narrow as for $A_1(\theta)$, even in 
the limit of $a\rightarrow 0$. 

\begin{figure}
\centerline{\includegraphics[width=0.5\textwidth]{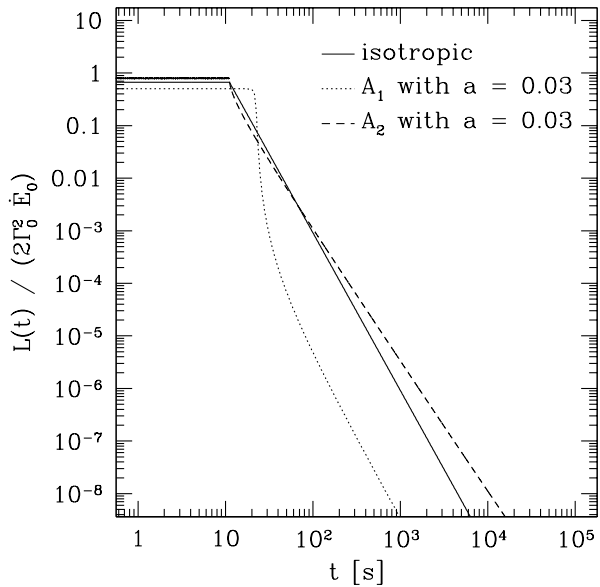}}
\caption{
Bolometric light curve from a shell whose emission suddenly cuts off
(see text).
}
\label{fig:lc_cut}
\end{figure}

As a simple illustration, consider a spherical thin shell with constant 
emission power $\dot{E}_0$, which is moving with $\Gamma_0=300$, and 
suppose that its emission suddenly cuts off at radius 
$R_\mathrm{cut}=6\times 10^{16}$~cm.
\reffig{fig:lc_cut} shows the produced bolometric light curve.
It depends on the intrinsic anisotropy of the source, $A(\theta)$.
We show three cases: isotropic emission $A(\theta)=1$, $A_1(\theta)$
and $A_2(\theta)$ (same as in Fig.~3). 
The limb-brightened emission $A_1(\theta)$ produces a very steep decay
  in the light curve.

%%%%%%%%%%%%%%%%%%%%%%%%%%%%%%%%%%%%%%%%%%%%%%%%%%%%%%%%%
%%%%%%%%%%%%%%%%%%%%%%%%%%%%%%%%%%%%%%%%%%%%%%%%%%%%%%%%%

We conclude that extremely fast variations in the light curve, e.g. short
flares or steep drops, may be observed in synchrotron afterglows
if electrons are preferentially accelerated along $\bB$, as the response 
function can be arbitrarily narrow for $a\ll 1$. Note that the emission 
is limb-brightened in this case, i.e. the situation is opposite to what 
was considered in section~3, and the description using $k>1$ does not apply. 
Limb-brightening has little or no effect on $\tjet$ or $R_\perp$, in 
contrast to the 
limb-darkened
model of section~3. 

Another implication of the preferential electron acceleration along $\bB$
is the reduction of synchrotron emissivity by a factor of $\sim a^2$
(cf. $\eta_1$ in \refeq{eq:eta12}). All synchrotron-emission formulae contain
only the component of $\bB$ perpendicular to the electron velocity, which
equals $\sim aB$ for $a\ll 1$. Then the effective $\epsilon_B$ that would
be inferred from the data using isotropic models will underestimate the
real $\epsilon_B$ by a factor $\sim a^{-2}$.

   The synchrotron model with
angular distribution $A_2(\theta)$, does not predict 
  significant changes in the afterglow light curve,
because limb-darkening is never 
  strong for synchrotron emission, regardless of $a$. 
A moderate change in $\tjet$ 
and a less pronounced jet break may be expected compared with the case of 
isotropic emission.

%%%%%%%%%%%%%%%%%%%%%%%%%%%%%%%%%%%%%%%%%%%%%%%%%%%%%%%%%%%%%%%%%%%%%%%%%%
%%%%%%%%%%%%%%%%%%%%%%%%%%%%%%%%%%%%%%%%%%%%%%%%%%%%%%%%%%%%%%%%%%%%%%%%%%
%%%% DISCUSSION
%%%%%%%%%%%%%%%%%%%%%%%%%%%%%%%%%%%%%%%%%%%%%%%%%%%%%%%%%%%%%%%%%%%%%%%%%%
%%%%%%%%%%%%%%%%%%%%%%%%%%%%%%%%%%%%%%%%%%%%%%%%%%%%%%%%%%%%%%%%%%%%%%%%%%
  
\section{Discussion}
\label{sec:discussion}

The usual assumption of isotropic emission in the rest frame of the blast 
wave is likely to be invalid. 
Even the standard synchrotron model with isotropic electron distribution 
  produces anisotropic, limb-darkened radiation (\refsec{sec:meca}). 
This fact is a consequence of the preferential orientation of the magnetic 
field in the blast wave.  Strong limb-brightening is also possible if the 
radiating electrons are preferentially accelerated along the magnetic field. 

Anisotropy may resolve a few puzzles encountered in afterglow modeling: \\
(i) The usual argument that the curvature effect filters out fast 
variability, prohibiting 
  strong variations 
in the light curve on timescales 
$\Delta t <\tau=R/2\Gamma^2 c$, is not valid for anisotropic emission. 
An anisotropic variable spherical source can produce fast changes in the 
light curve, similar to observed bizarre features in GRB afterglows. 
This result holds for both limb-darkened and limb-brightened types of 
anisotropy. It suggests that the X-ray flares observed by {\it Swift} 
with $\Delta t/\tobs\la 0.1$ do not necessarily imply an additional 
component of internal origin. Instead they may be produced, e.g. by the 
reverse shock in the blast wave, whose emission may suddenly brighten 
and weaken as the reverse shock propagates into the inhomogeneous ejecta 
of the explosion. This model may also explain sudden steep drops in the 
afterglow light curve as observed in GRB~070110 (Troja et al. 2008).
This explanation assumes
that the X-ray radiating particles are cooling 
fast compared with the jet expansion timescale, as slow cooling would 
suppress short time-scale variations of the source luminosity.

\begin{figure}
\centerline{\includegraphics[width=0.5\textwidth]{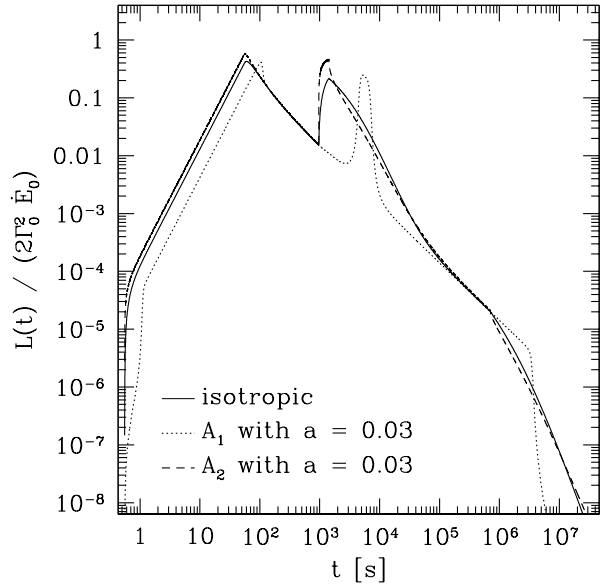}}
\caption{
Bolometric light curve for a toy afterglow model (see text).
The result is plotted for three cases: isotropic emission in the rest frame 
of the blast wave $A(\theta)=1$ (solid curve), 
limb-brightened synchrotron emission $A_1(\theta)$ with $a=0.03$ 
(dotted curve; see Section~4 for the description of the synchrotron model)
and limb-darkened synchrotron emission $A_2(\theta)$ with $a=0.03$ 
(dashed curve).
}
\label{fig:lc}
\end{figure}

Examples of such short times-cale behaviors are given by the toy model 
in \reffig{fig:lc}. It shows the 
  synchrotron emission 
produced by a thin shell with Lorentz factor 
$\Gamma(R)=\Gamma_0=300$ at 
$R<R_\mathrm{dec}=3\times 10^{17}\,\mathrm{cm}$
and $\Gamma(R)=\Gamma_0\left(R/R_\mathrm{dec}\right)^{-3/2}$ at 
$R>R_\mathrm{dec}$. This approximately describes a blast wave decelerating
in a uniform medium. The shell is assumed to radiate with bolometric power 
proportional to the dissipation rate in the blast wave, which gives
$\dot{E}(R)=\dot{E}_0\left(R/R_\mathrm{dec}\right)^2$ at $R<R_\mathrm{dec}$
and $\dot{E}(R)=\dot{E}_0\left(R/R_\mathrm{dec}\right)^{-1}$ at 
$R>R_\mathrm{dec}$. A realistic blast wave has two shocks -- forward and 
reverse -- and both can produce 
a long-lived afterglow. Our toy model may describe the emission from 
either shock, although it is very much simplified. 
To illustrate the curvature effect on variability, we add two features:
a sudden brief increase in $\dot{E}(R)$ at $R=3\,R_\mathrm{dec}$ (which 
simulates a flare) and the abrupt cutoff of $\dot{E}(R)$ at 
$15\,R_\mathrm{dec}$. 
For comparison, we show the light curves produced for 
  three cases: isotropic emission in the rest frame of the shell,
  limb-brightened emission and limb-darkened emission described in Section~4.

%%%%%%%%%%%%%%%%%%%%%%%%%%%%%%%%%%%%%%%%%%%%%%%%%%%%%%%%%
%%%%%%%%%%%%%%%%%%%%%%%%%%%%%%%%%%%%%%%%%%%%%%%%%%%%%%%%%

(ii) If a relativistic source is limb-darkened, most of its 
emission in the fixed lab frame is confined within an angle {\it smaller} 
than $\Gamma^{-1}$. This effect suggests a 
possible explanation for 
the lack of jet-break detections in GRBs, as the increased beaming 
significantly delays the jet break in the observed light curve (Section~3.2). 
We also discussed in Section~3.3 the consequences of such anisotropy for
the apparent size of the radio afterglow source. Although the strong
limb-darkening appears to be impossible for standard synchrotron 
afterglows, it may be possible for a different radiative mechanism.
For example, 
limb-darkening may be expected for the jitter mechanism (Medvedev \& Loeb 
1999), as the electrons are preferentially accelerated perpendicular to 
the shock plane and radiate preferentially in the radial direction. 
 
While this paper was focused on afterglow, the source of prompt GRB 
emission may also be intrinsically anisotropic. This may 
impact
models that propose the curvature effect to control the steep 
X-ray decay at the end of the prompt emission (see e.g. Genet \& Granot 2009; 
Zhang et al. 2009). The effect can be seen in Figure~4. 
Anisotropy of 
 the
prompt emission may also change
the optical depth of the source to high-energy photons, $\tau_{\gamma\gamma}$,
as the cross section for $\gamma\gamma$ reaction strongly depends 
on the angle between photons. This may affect the constraints on the 
Lorentz factor of the jet that are inferred from $\tau_{\gamma\gamma}<1$.
The effect is especially strong for emission without front-back
  symmetry in the source frame; such asymmetric emission would be 
  a more radical assumption compared with the ordinary limb-brightening or 
  limb-darkening considered in this paper.

\section*{Acknowledgments}
 We thank Jonathan Granot and the referee for comments on the manuscript.
AMB was supported by NASA Swift grant, Cycle 4. 
ZLU is supported by WCU program (R32-2009-000-10130-0) of NRF/MEST of Korea.
FD and RM were supported by the 
  French
space agency (CNES).

%####################################################################


\begin{thebibliography}{}
\bibitem{}
Burrows D.N., Racusin J., 2006, NCimB, 121, 1273
\bibitem{}
Burrows D.N. et al., 2005, Sci, 309, 1833 
\bibitem{}
Chincarini G. et al., 2007, ApJ, 671, 1903
\bibitem{}
Covino S. et al., 1999, A\&A, 348, L1
\bibitem{}
Frail D.A., 1997, Nature, 389, 261
\bibitem{}
Genet F., Daigne F., Mochkovitch R., 2007, MNRAS, 381, 732   
\bibitem{}
Genet F. \& Granot J., 2009, MNRAS, 399, 1328   
\bibitem{}
Granot J., Piran, T., Sari, R., 1999, ApJ, 527, 236
\bibitem{}
Grupe D. et al., 2007, ApJ, 662, 443
\bibitem{}
Hededal, C.B. et al., 2004, ApJ 617, L107
\bibitem{}
Ioka K., Kobayashi S., Zhang B., 2005, ApJ, 631, 429 
\bibitem{}
Kumar P., Narayan R., Johnson J.L., 2008, MNRAS, 388, 1729
\bibitem{}
Kumar P., Panaitescu, A., 2000, ApJ, 541, L51 
\bibitem{}
Lazzati D., Perna R., 2007, MNRAS, 375, L46
\bibitem{}
Lyutikov, M., 2006, MNRAS, 369, L5
\bibitem{}
Nishikawa, K.-I. et al., 2009, ApJ 698, L10
\bibitem{}
Nousek J.A. et al., 2006, ApJ, 642, 389
\bibitem{}
Oren Y., Nakar E., Piran T., 2004, MNRAS, 353, L35
\bibitem{}
Pihlstr\"om Y.M., Taylor G.B., Granot J., Doeleman S., 2007, ApJ, 664, 411   
\bibitem{}
Rhoads J., 1997, ApJ 487, L1 
\bibitem{}
Sato G. et al., 2007, ApJ, 657, 359
\bibitem{}
Shady P. et al., 2007, MNRAS, 380, 1041
\bibitem{}
Spitkovski, A., 2008, ApJ 682, L5
\bibitem{}
Taylor G.B. et al., 2005, ApJ, 622, 986 
\bibitem{}
Troja E. et al., 2007, ApJ, 665, 599
\bibitem{}
Uhm Z.L., Beloborodov A.M., 2007, ApJ, 665, L93
\bibitem{}
Willingale R. et al., 2007, ApJ, 662, 1093 
\bibitem{}
Zhang B. et al., 2006, ApJ, 642, 354
\bibitem{}
Zhang B.-B. et al., 2009, ApJ, 690, L10

\end{thebibliography}
\end{document}